\documentclass[aps,prl,showkeys,floatfix,twocolumn,bibnotes,nofootinbib,superscriptaddress]{revtex4-2}
\usepackage{graphicx,amsmath,amssymb,hyperref,natbib,slashed}
\usepackage[dvipsnames]{xcolor}
\usepackage{xspace}
\usepackage[normalem]{ulem}
\usepackage[utf8]{inputenc}
\usepackage[T1]{fontenc}

\renewcommand{\d}{\mathrm{d}}
\newcommand{\ul}[1]{%
	\underline{\smash{#1}\vphantom{T}}\vphantom{#1}%
}
\newcommand{\eqsp}{\,}
\newcommand{\be}{\begin{equation}}
\newcommand{\ee}{\end{equation}}

\begin{document}
\hfill CERN-TH-2022-099, DESY 22-102, ULB-TH/22-09

\title{Minimal sterile neutrino dark matter}

\newcommand{\AddrOslo}{%
Department of Physics, University of Oslo, Box 1048, N-0316 Oslo, Norway}
\newcommand{\AddrDESY}{%
Deutsches Elektronen-Synchrotron DESY, Notkestr.~85, 22607 Hamburg, Germany}
\newcommand{\AddrBrussels}{%
Service de Physique Théorique, Université Libre de Bruxelles, Boulevard du Triomphe, CP225, B-1050 Brussels, Belgium}
\newcommand{\AddrNYU}{%
Center for Cosmology and Particle Physics, Department of Physics, New York University, New York, NY 10003, USA}
\newcommand{\AddrKITP}{%
Kavli Institute for Theoretical Physics, University of California, Santa Barbara, CA 93106, USA}
\newcommand{\AddTAU}{%
School of Physics and Astronomy, Tel-Aviv University, Tel-Aviv 69978, Israel\vspace*{0.1cm}}
\newcommand{\AddrUiB}{%
Department of Physics and Technology, University of Bergen, 5020 Bergen, Norway}
\newcommand{\AddrKIAS}{%
Korea Institute for Advanced Study, Seoul 02455, Republic of Korea}
\newcommand{\AddrCERN}{%
Theoretical Physics Department, CERN, 1211 Geneva 23, Switzerland}
\newcommand{\AddrMPIK}{%
Max-Planck-Institut f\"ur Kernphysik, Saupfercheckweg 1, 69117 Heidelberg, Germany}

\author{Torsten Bringmann}
\email{torsten.bringmann@fys.uio.no}
\affiliation{\AddrOslo}
\affiliation{\AddrCERN}

\author{Paul Frederik Depta}
\email{frederik.depta@mpi-hd.mpg.de}
\affiliation{\AddrMPIK}

\author{Marco Hufnagel}
\email{marco.hufnagel@ulb.be}
\affiliation{\AddrBrussels}

\author{J\"orn Kersten}
\email{joern.kersten@uib.no}
\affiliation{\AddrKIAS}
\affiliation{\AddrUiB}

\author{Joshua T.\ Ruderman}
\email{ruderman@nyu.edu}
\affiliation{\AddrNYU}

\author{Kai Schmidt-Hoberg}
\email{kai.schmidt-hoberg@desy.de}
\affiliation{\AddrDESY}

\begin{abstract}
We propose a novel mechanism to generate  sterile neutrinos $\nu_s$ in the early Universe,
by converting ordinary neutrinos $\nu_\alpha$ in scattering processes $\nu_s\nu_\alpha\to\nu_s\nu_s$.
After initial production by oscillations, this leads to an exponential growth in the $\nu_s$
abundance.
We show that such a production regime naturally occurs for self-interacting $\nu_s$,
and that this opens up significant new parameter space where  $\nu_s$
make up all of the observed dark matter. Our results provide strong motivation to further push
the sensitivity of X-ray line searches, and to improve on constraints from structure formation.
\end{abstract}

\date{April 26, 2023}

\maketitle

\paragraph*{Introduction.---}%
The existence of sterile neutrinos, putative particles that are uncharged under the standard model (SM) gauge interactions, 
is extremely well motivated.
For example, such sterile states provide 
natural candidates~\cite{Minkowski:1977sc,Yanagida:1980,Glashow:1979vf,Gell-Mann:1980vs,Mohapatra:1980ia} to explain the 
observed tiny nonzero neutrino masses~\cite{SajjadAthar:2021prg}.
If one of the sterile neutrinos has a mass in the keV range and is stable
on cosmological time scales, furthermore, it is an excellent candidate for the dark matter (DM)
in our Universe~\cite{Boyarsky:2018tvu}. A smoking-gun signal for this scenario would
be an astrophysical X-ray line, resulting from DM decaying into an active neutrino and a
photon~\cite{Abazajian:2001vt}. Such X-ray signatures are very actively being searched for, leading to ever more stringent
limits on how much sterile and active neutrinos can mix~\cite{Gerbino:2022nvz}
(while Refs.~\cite{Bulbul:2014sua,Boyarsky:2014jta} report a potential detection).

Sterile neutrinos can be produced by neutrino oscillations in the early
Universe, which is known as the Dodelson-Widrow (DW) mechanism~\cite{Dodelson:1993je}.
However, the region of parameter space where this
mechanism produces an abundance of sterile neutrinos consistent with
the entirety of the observed DM is excluded~\cite{Abazajian:2017tcc}. 
Alternative scenarios that remain viable include resonant production
in the presence of a large lepton asymmetry~\cite{Shi:1998km}, production
by the decay of a
scalar~\cite{Shaposhnikov:2006xi,Kusenko:2006rh,Petraki:2007gq,Roland:2014vba,Merle:2015oja,Konig:2016dzg},
via an extended gauge sector~\cite{Bezrukov:2009th,Kusenko:2010ik,Dror:2020jzy},
and production by oscillations modified by new self-interactions of the SM
neutrinos~\cite{DeGouvea:2019wpf,Kelly:2020pcy,Chichiri:2021wvw,Benso:2021hhh}
or by interactions between the sterile neutrinos and a significantly heavier
scalar 
in combination with a large lepton asymmetry~\cite{Hansen:2017rxr}.

All known viable mechanisms thus require new particles in addition
to the sterile neutrino, either explicitly or implicitly (e.g., to
ensure gauge invariance or to create a lepton asymmetry much larger than
the observed baryon asymmetry).  Our objective here is to propose a novel scenario
that is minimal in the sense that it requires only a \textit{single} real
degree of freedom on top of the DM candidate.

Recently, some of us introduced a new generic DM production 
mechanism~\cite{Bringmann:2021tjr} that is characterized by
DM particles transforming heat bath particles into more DM, thereby triggering an era of exponential growth of the DM
abundance (see also Ref.~\cite{Hryczuk:2021qtz}). As discussed there, 
one possibility to provide the necessary DM seed population
for such a DM density evolution is through an initial 
freeze-in~\cite{Hall:2009bx} period.
In the scenario proposed in this article,
instead, sterile neutrinos $\nu_s$ are initially generated through oscillations like in the DW mechanism
and subsequently transform active neutrinos $\nu_\alpha$ through 
the process $\nu_s\nu_\alpha\to\nu_s\nu_s$.
We demonstrate that such a scenario generically emerges when sterile neutrinos feel the presence of a dark force,
and that this opens up significant portions of parameter space for sterile neutrino DM that may be detectable
with upcoming experiments.

\smallskip
\paragraph*{Model setup.---}%
A necessary requirement to realize DM production via exponential growth is
$\langle\sigma v\rangle_{\rm tr}\gg \langle\sigma v\rangle_{\rm fi}$~\cite{Bringmann:2021tjr}, where
$\langle\sigma v\rangle_{\rm tr}$ is the thermally averaged interaction rate for the transmission process,
i.e.~the conversion of a heat bath particle to a DM particle, and $\langle\sigma v\rangle_{\rm fi}$ is the corresponding
quantity for the more traditional freeze-in process~\cite{Hall:2009bx}, where a pair of DM particles is produced from the 
collision of heat bath particles. 
Here we point out that a simple and generic way to realize this condition is a secluded dark
sector~\cite{Pospelov:2007mp,Feng:2008mu,Pospelov:2008zw,Cheung:2010gj}
where DM particles interact among each other via some mediator $\phi$, while interacting with the visible sector only through
mixing by an angle~$\theta$. In that case, both processes dominantly proceed via the $s$-channel
exchange of $\phi$, resulting in $\langle\sigma v\rangle_{\rm tr}\propto \sin^2\theta$
and $\langle\sigma v\rangle_{\rm fi}\propto \sin^4\theta$.
We note that such `secret interactions' of sterile neutrinos
have been studied in different cosmological contexts
before~\cite{Hannestad:2013ana,Dasgupta:2013era,Bringmann:2013vra,Ko:2014bka,Archidiacono:2014nda,Mirizzi:2014ama,Tang:2014yla,Saviano:2014esa,Kouvaris:2014uoa,Chu:2015ipa,Archidiacono:2015oma,Tabrizi:2015bba,Binder:2016pnr,Archidiacono:2016kkh,Forastieri:2017oma,Bezrukov:2017ike,Jeong:2018yts,Song:2018zyl,Chu:2018gxk,Blennow:2019fhy,Ballett:2019pyw,Johns:2019cwc,Pires:2019elj,Archidiacono:2020yey,Berbig:2020wve,Corona:2021qxl}.

Motivated by these general considerations, we concentrate in the following on
a single sterile neutrino $\nu_s$ interacting with a light scalar $\phi$,
both singlets under the SM gauge group.
Assuming Majorana masses for $\nu_s$ and the active neutrinos, $\nu_\alpha$, the relevant
Lagrangian terms are given by
\begin{equation}
\mathcal{L}\supset
-\frac{1}{2}\overline{\nu_s^c} m_s \nu_s
-\overline{\nu_\alpha} m_{\alpha s} \nu_s
-\frac{1}{2}\overline{\nu_\alpha} m_\alpha \nu_\alpha^c
+ \frac{y}{2} \overline{\nu_s^c} \phi \nu_s
+ \text{h.c.}, \nonumber
\label{eq:L1}
\end{equation}
where repeated indices $\alpha$ are summed over and
$\nu_\alpha$\,($\nu_s$) are left-\,(right-)\,handed spinors. 
We will concentrate on the case of heavy mediators, $m_\phi>2\, m_s$, for most of this article,
but later also briefly discuss phenomenological consequences of lighter mediators.
We checked that even for large mediator self-couplings, number-changing interactions like $2 \phi \leftrightarrow 4 \phi$ or $3 \phi \leftrightarrow 4 \phi$ do not qualitatively change our results and therefore neglect the scalar potential in all practical calculations.
We further assume, for simplicity and concreteness, that $\nu_s$
dominantly mixes only with the active neutrino species $\nu_e$, and that $m_s\gg m_\alpha$.
Expressed in terms of mass eigenstates, which for ease of notation we denote by the same symbols as flavor eigenstates, 
the interactions of the mediator then take the form
\begin{equation}
\mathcal{L}_\phi^\text{I}=
\frac{y}{2}\phi\, \big(
\cos^2\!\theta\, \overline{\nu_s^c} \nu_s
- \sin(2\theta)\, \overline{\nu_\alpha} \nu_s
+ \sin^2\!\theta\, \overline{\nu_\alpha} \nu_\alpha^c
\big)
+ \text{h.c.}
\label{eq:phicouplings}
\end{equation}
with $\theta \simeq m_{\alpha s}/m_s\ll1$.
The unsuppressed couplings among $\phi$ and $\nu_s$ turn out to be sufficiently strong to
equilibrate the dark sector during the new exponential production phase that
we consider below.
On the other hand, mass-mixing-induced interactions between $\nu_s$ and electroweak gauge bosons
 are suppressed by the Fermi constant, $G_F$,  and will only be relevant in setting the initial sterile neutrino abundance.

\begin{figure}[t]
\begin{picture}(100, 45)
\put(-70,15){\includegraphics[width=0.32\columnwidth]{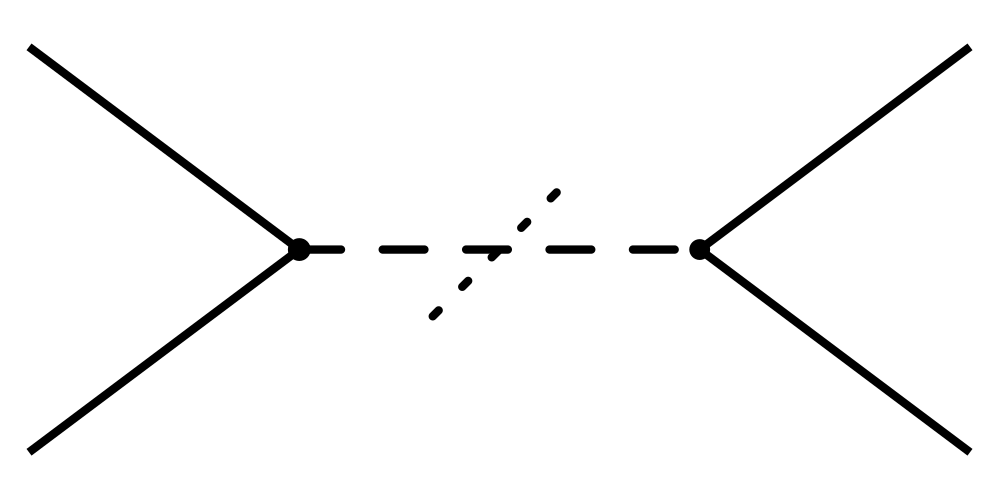}}
\put(10,15){\includegraphics[width=0.32\columnwidth]{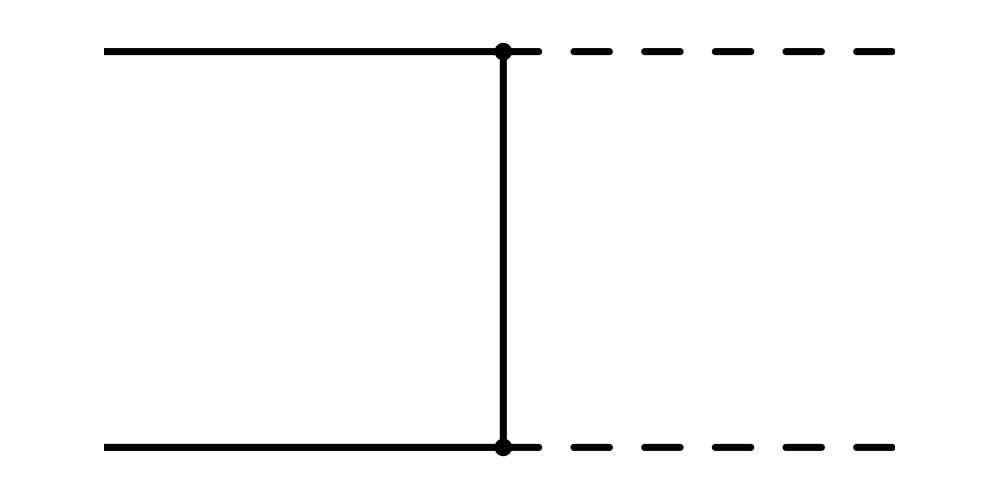}}
\put(90,15){\includegraphics[width=0.32\columnwidth]{figures/diagram1.png}}

\put(-60, 15){$\nu_s$}
\put(-60, 50){$\nu_s$}
\put(-10, 50){$\nu_s$}
\put(-10, 15){$\nu_s$}
\put(-34, 22){$\phi$}

\put(20, 23){$\nu_s$}
\put(20, 43){$\nu_s$}
\put(38, 33){$\nu_s$}
\put(73, 25){$\phi$}
\put(73, 41){$\phi$}

\put(100, 15){$\nu_\alpha$}
\put(100, 50){$\nu_s$}
\put(150, 50){$\nu_s$}
\put(150, 15){$\nu_s$}
\put(126, 22){$\phi$}

\put(-37, 0){(a)}
\put(44, 0){(b)}
\put(124, 0){(c)}
\end{picture}
\caption{%
Relevant diagrams for (a) \emph{dark sector thermalization}, $\nu_s \nu_s \!\to\! \phi^* \!\!\to\! \nu_s \nu_s$, 
(b) \emph{additionally increasing the dark sector number density}
between DW production and exponential growth,
$\nu_s \nu_s \to \phi \phi$, and (c) \emph{exponential growth of DM}, $\nu_s \nu_\alpha  \!\to\! \phi^* \!\!\to\!  \nu_s \nu_s$.
Since $\phi$ is (almost) on-shell for (a) and (c),
it is sufficient to include only the rates for $\nu_s \nu_s \leftrightarrow \phi$ and $\nu_s \nu_\alpha \to \phi$. 
See text for further details.
}
\label{fig:diagrams}
\end{figure}

\smallskip
\paragraph*{Evolution of $\nu_s$ number density.---}%
For an initially vanishing abundance, in particular, the interactions in Eq.~(\ref{eq:phicouplings}) only allow freeze-in production
of $\nu_s$. While the corresponding rate scales as $\sin^4\theta$,  active-sterile neutrino oscillations at temperatures
above and around $\Lambda_{\rm QCD}\sim150$\,MeV, in combination with neutral and charged current interactions
with the SM plasma, lead to a production rate scaling as
$\sin^2\theta$~\cite{Dodelson:1993je}. Adopting results from Ref.~\cite{Asaka:2006nq}, we 
use the $\nu_s$ number density, $n_s$, and
energy density, $\rho_s\sim\langle p\rangle n_s$, 
that result from this DW production. Once it is completed, and in the
absence of dark sector interactions,
the expansion of the Universe will decrease these quantities
as $n_s \propto a^{-3}$ and $\rho_s\propto a^{-4}$, respectively,
where $a$ is the scale factor.

Some time later, various decay and scattering processes (cf.\ Fig.~\ref{fig:diagrams}) become relevant
due to the new interactions appearing in Eq.~(\ref{eq:phicouplings}) and,
for the parameter space we are interested in here, eventually thermalize the dark sector
via the (inverse) decays $\nu_s \nu_s \leftrightarrow \phi$.
From that point on, the phase-space densities of $\nu_s$ and $\phi$ follow Fermi-Dirac and Bose-Einstein
distributions, respectively, that are described by a common
dark sector temperature $T_\mathrm{d}$ as well as chemical potentials $\mu_s$ and $\mu_\phi$. Similar to the
situation of freeze-out in a dark sector~\cite{Bringmann:2020mgx}, the evolution of these quantities is determined by
a set of Boltzmann equations for the number densities $n_{s,\phi}$ and total dark sector energy density $\rho=\rho_\phi + \rho_s$:
\begin{align}
\dot{n}_s + 3 H n_s &= C_{n_s}\,, \label{eq:Boltz_nd} \\
\dot{n}_\phi + 3 H n_\phi &= C_{n_\phi}\,, \label{eq:Boltz_nphi} \\
\dot{\rho} + 3 H (\rho + P) &= C_\rho\,, \label{eq:Boltz_rhodphi}
\end{align}
where $H\equiv\dot a/a$ is the Hubble rate, $P = P_s + P_\phi$ is the total dark sector
pressure, and $C_i$ are the various collision operators
(see Appendix for details).
With  $\phi \leftrightarrow \nu_s \nu_s$ in equilibrium, the chemical potentials are related by $2 \mu_s = \mu_\phi$,
allowing us to
replace the first two of the above equations with a single differential equation for $\tilde n \equiv n_s + 2 n_\phi$.
Noting that $\rho\propto a^{-4}$ and $\tilde n \propto a^{-3}$, both right before
and after $\phi \leftrightarrow \nu_s \nu_s$ starts to dominate over the Hubble rate,
the initial conditions to the coupled differential equations for $\tilde n$ and $\rho$
can then be determined at the end of DW production.

\begin{table}
\centering
\begin{tabular}{c||c|c|c|c}
&$m_s$ & $m_\phi$ & $\sin^2 (2 \theta)$  & $y$ \\
\hline
\hline
\emph{BP1} & 12\,keV & 36\,keV & $2.5 \times 10^{-13}$  & $1.905 \times 10^{-4}$ \\
\hline
\emph{BP2} & 20\,keV & 60\,keV & $3.0 \times 10^{-15}$  & $1.602 \times 10^{-3}$ \\
\end{tabular}
\caption{Parameter values for the two benchmark points considered in Fig.~\ref{fig:evo}.
}
\label{tab:BP}
\end{table}

\begin{figure*}[t]
\includegraphics[width=\columnwidth]{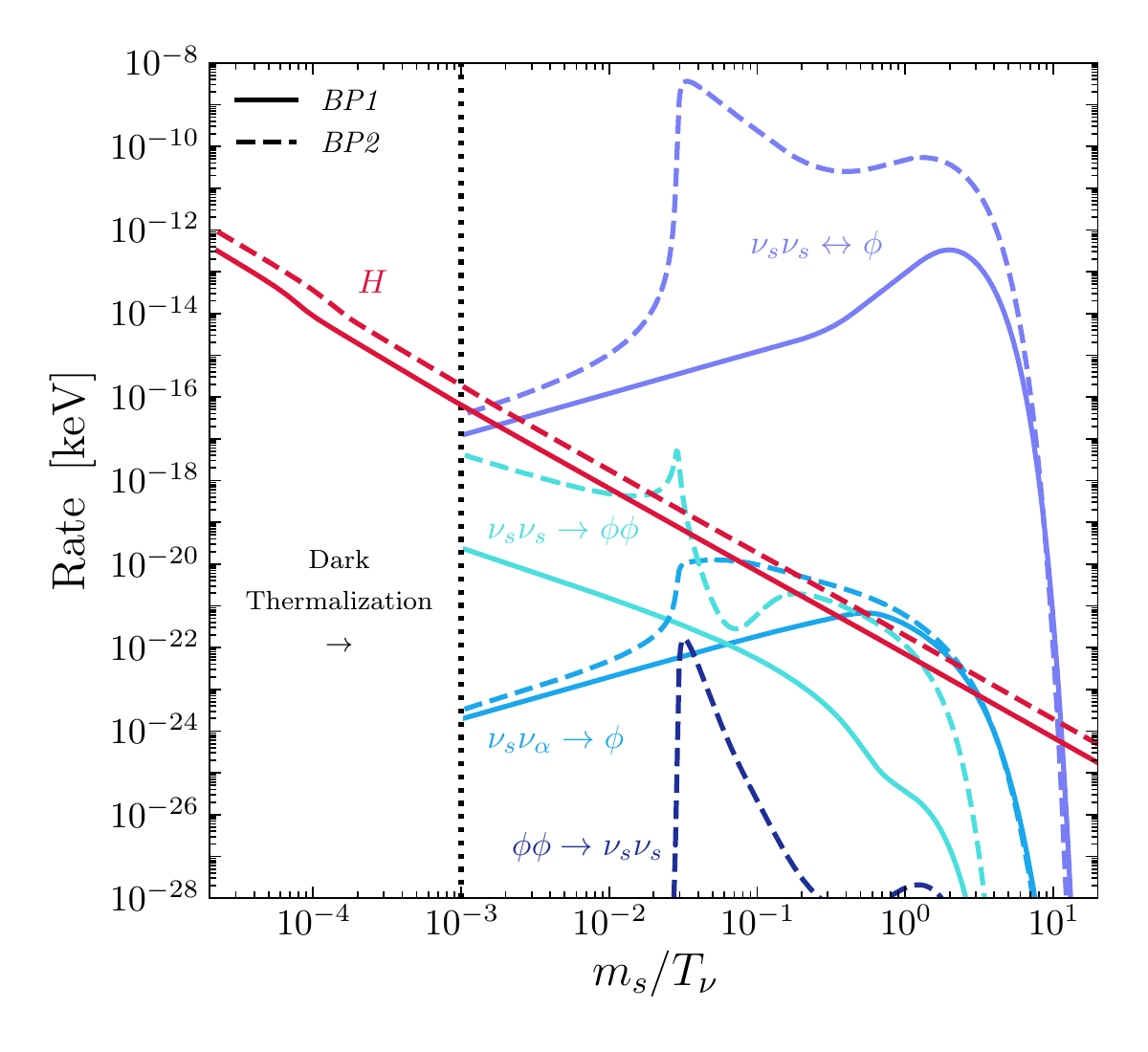}
\includegraphics[width=\columnwidth]{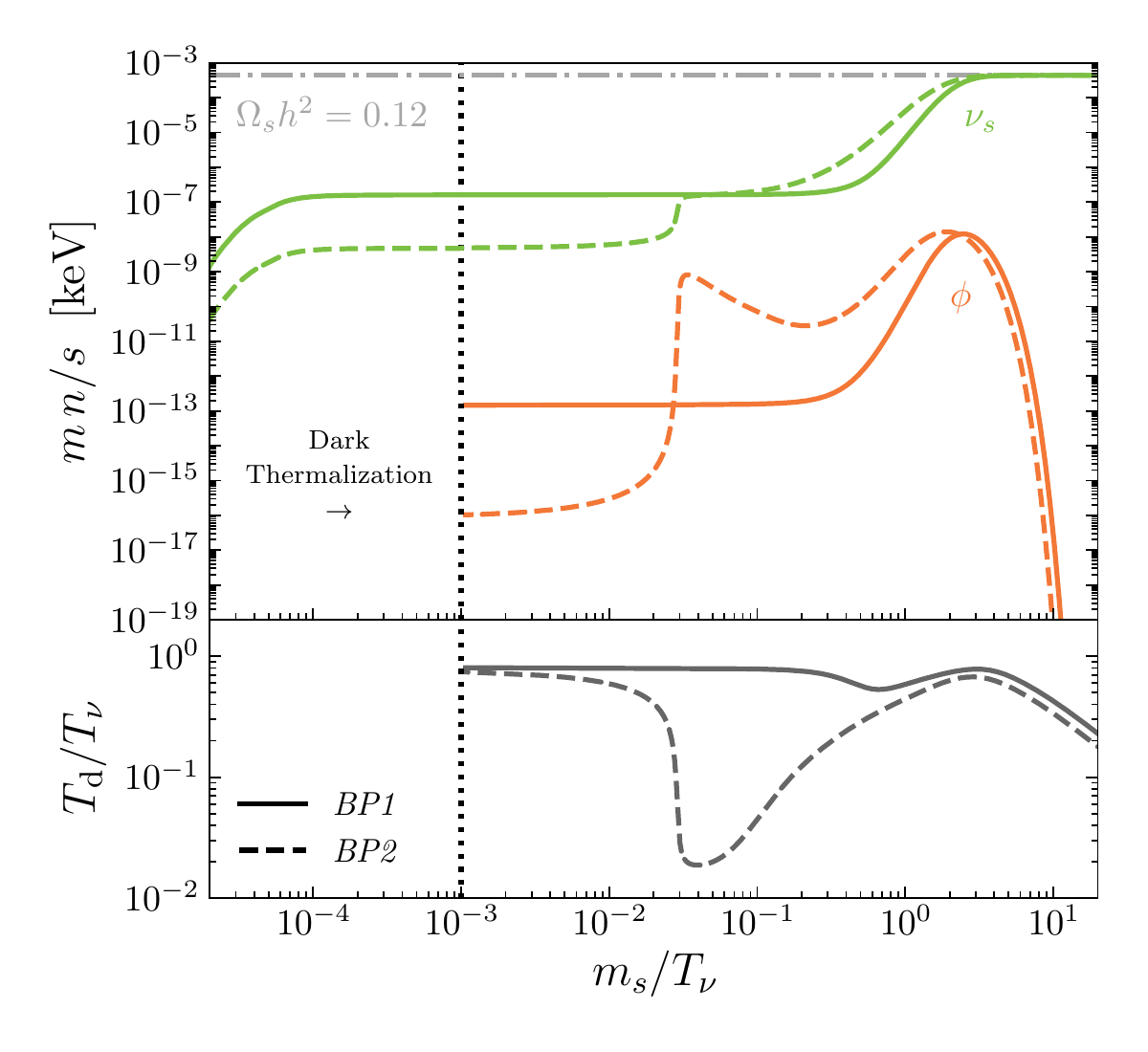}
\caption{%
Cosmological evolution for the benchmark points \emph{BP1} (solid lines) and  \emph{BP2} (dashed lines)
specified in Tab.~\ref{tab:BP}, as function of the inverse active neutrino temperature.
{\it Left}: Comparison of Hubble rate $H$ (red) with the contribution of the indicated processes to $\left|C_{n_s}\right|/n_s$ 
(blue lines), see Appendix for details. 
{\it Right}: Corresponding evolution of $\nu_s$ (green) and $\phi$
(orange) abundances, and temperature ratio (dark gray, bottom panel).
The dash-dotted gray line indicates the observed DM abundance~\cite{Planck:2018vyg}.
}
\label{fig:evo}
\end{figure*}

In order to illustrate  the subsequent evolution of the system, let us consider two concrete benchmark points,
cf.~Tab.~\ref{tab:BP}, for which the sterile neutrinos obtain a relic density that matches the observed DM abundance
of $\Omega_{\rm DM} h^2\simeq0.12$~\cite{Planck:2018vyg}, with a mixing angle too small to achieve this with standard
DW production. 
As demonstrated in Fig.~\ref{fig:evo}, with solid (dashed) lines for \emph{BP1} (\emph{BP2}), this 
leads to qualitatively different behaviors:
\begin{itemize}
\item[\emph{BP1}] Here the only additional process (beyond $\phi \leftrightarrow \nu_s \nu_s$) where 
the rate becomes comparable to $H$, at $m_s / T_\nu \sim 0.2$ with $T_\nu$ the active neutrino temperature,
is $\nu_s \nu_\alpha\to \phi$ (left panel, light blue).  This triggers exponential growth in the abundance for both $\nu_s$ and $\phi$
(right panel, green and orange) through $\nu_s \nu_\alpha \to \phi^* \to \nu_s \nu_s$,
with $\phi$ being (almost) on shell, cf.\ Fig.~\ref{fig:diagrams}~(c).
Once $T_\nu \ll m_\phi$ the transmission process becomes inefficient and the final $\nu_s$ abundance is obtained.
Afterwards, since both $\phi$ and $\nu_s$ are non-relativistic, the dark sector
temperature decreases with $T_\mathrm{d}\propto a^{-2}$ both before and after kinetic decoupling (right panel, dark gray).

\item[\emph{BP2}] In this case the larger coupling $y$  (needed to compensate for the smaller $\theta$) leads to another
process impacting the evolution of the system: at $m_s / T_\nu \sim 0.01$, the rate for $\nu_s \nu_s \to \phi \phi$
(left panel, cyan, and Fig.~\ref{fig:diagrams}~(b) starts to be comparable to $H$. As $\phi$ predominantly decays into $\nu_s \nu_s$ (left panel, orchid),
this effectively transforms kinetic energy to rest mass by turning $2 \nu_s$ to $4 \nu_s$ -- very similar
to the reproductive freeze-in mechanism described by Refs.~\cite{Hansen:2017rxr,Mondino:2020lsc,March-Russell:2020nun}. As expected, this leads to
a significant drop in the temperature $T_\mathrm{d}$ (right panel, black).
This process becomes inefficient  for $T_\mathrm{d} \ll m_\phi$, due to the Boltzmann suppression of $\phi$.
Subsequently, the rate for $\nu_s \nu_\alpha \to \phi$ (left panel,
light blue) becomes comparable to $H$,
leading to a phase of exponential growth in the same way as for \emph{BP1}.
\end{itemize}

\smallskip
\paragraph*{Observational constraints.---}%
Due to the mixing with active neutrinos, $\nu_s$ is not completely stable and subject to the same decays as in
the standard scenario for keV-mass sterile neutrino DM\@.
The strongest constraints on these decays come from a variety of X-ray line searches. We take the compilation of limits
from Ref.~\cite{Gerbino:2022nvz}, but only consider the overall envelope of constraints from 
Refs.~\cite{Horiuchi:2013noa,Malyshev:2014xqa,Foster:2021ngm,Sicilian:2020glg,Roach:2019ctw,Boyarsky:2007ge}.
Furthermore,
we consider projections for eROSITA~\cite{Dekker:2021bos}, Athena~\cite{Ando:2021fhj}, and eXTP~\cite{Malyshev:2020hcc}.

Observations of the Lyman-$\alpha$ forest using light from distant quasars place stringent limits on a potential
cutoff in the matter-power spectrum at small scales, where the scale of this cutoff is related to
the time of kinetic decoupling, $t_\mathrm{kd}$. In our scenario, this is determined by DM self-interactions and
we estimate $t_\mathrm{kd}$ from $H n_s = C_{\nu_s \nu_s \to \nu_s \nu_s}$~\cite{Hryczuk:2022gay}, where
the collision term $C_{\nu_s \nu_s \to \nu_s \nu_s}$ is stated in the Appendix.
A full evaluation of \mbox{Lyman-$\alpha$} limits would require evolving
cosmological perturbations into the non-linear regime, which is beyond the scope of this work. Instead, we recast
existing limits on the two main mechanisms that generate such a cutoff.
At times $t<t_\mathrm{kd}$, DM self-scatterings prevent
overdensities from growing on scales below the sound horizon
$r_\mathrm{s} = \int_0^{t_\mathrm{kd}} \d t\, c_\mathrm{s} / a$,
where $c_\mathrm{s} = \sqrt{\d P / \d \rho}$ is the speed of sound in the dark sector~\cite{Egana-Ugrinovic:2021gnu}.
We use the results from Ref.~\cite{Vogelsberger:2015gpr} for cold DM in kinetic equilibrium with dark radiation
(with $c_\mathrm{s} = 1/\sqrt{3}$) to recast the current Lyman-$\alpha$ constraint on the mass of a warm DM (WDM)
thermal relic $m_{\rm WDM}>1.9 \, \mathrm{keV}$~\cite{Garzilli:2019qki} to the bound
$r_\mathrm{s} < 0.34 \, \mathrm{Mpc}$. 
Overdensities are also washed out by the free streaming of DM {\it after} decoupling.
We evaluate the free-streaming length as
$\lambda_\mathrm{fs} = \int_{t_\mathrm{kd}}^{t_{\rm nl}} \mathrm{d} t\, \langle v \rangle / a$, where 
$\langle v \rangle = \langle p/E \rangle$ is the thermally averaged DM velocity
 and we integrate up to times $t_{\rm nl}$ where structure formation becomes relevant at redshifts of roughly $z\sim50$.
 We translate the WDM constraint of $m_{\rm WDM}>1.9 \, \mathrm{keV}$ 
 to $\lambda_\mathrm{fs} < 0.24 \, \mathrm{Mpc}$, which we will apply in the following to our scenario. 
We note that the WDM bounds from Ref.~\cite{Garzilli:2019qki} are based on marginalizing 
over different reionization histories. Fixed reionization models 
tend to produce less conservative constraints, which however illustrate the future potential of Lyman-$\alpha$ probes
once systematic errors are further reduced 
($m_\mathrm{WDM} > 5.3 \, \mathrm{keV}$~\cite{Palanque-Delabrouille:2019iyz}, e.g., corresponds to 
$r_\mathrm{s} < 0.09 \, \mathrm{Mpc}$ and $\lambda_\mathrm{fs} < 0.07 \, \mathrm{Mpc}$, respectively).

DM self-interactions are also constrained by a variety of astrophysical observations at late times~\cite{Tulin:2017ara}.
We adopt $\sigma_T / m_s < 1\,\mathrm{cm^2/g}$ as a rather conservative limit, where $\sigma_T$ is the momentum
transfer cross-section as defined in Ref.~\cite{Kahlhoefer:2013dca}. 
Far away from the $s$-channel resonance, 
we find
$\sigma_T / m_s \simeq y^4 m_s / (4 \pi m_\phi^4) + \mathcal{O} (v^2)$, largely independent of the DM velocity
$v$. For such cross sections  cluster observations~\cite{Kaplinghat:2015aga,Andrade:2020lqq}, 
or the combination of halo surface densities over a large mass range~\cite{Bondarenko:2017rfu},
can be (at least) one order of magnitude more competitive than our reference limit of 
$\sigma_T / m_s < 1\,\mathrm{cm^2/g}$.

\smallskip
\paragraph*{Viable parameter space for sterile neutrino DM\@.---}%

\begin{figure}[t]
\includegraphics[width=\columnwidth]{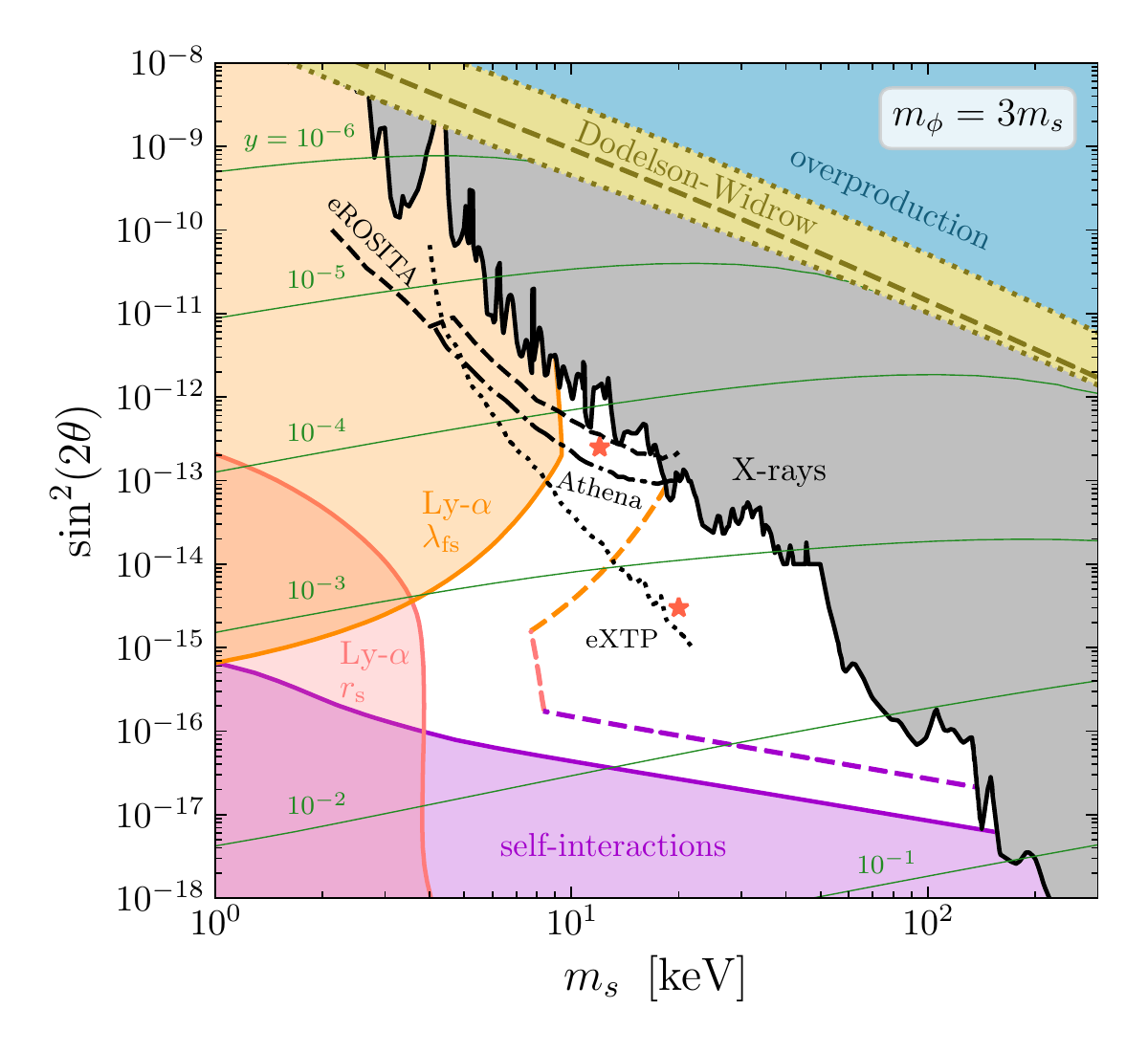}
\caption{%
Available parameter space in the $\sin^2(2\theta) - m_s$ plane, for $m_\phi=3m_s$.
The Yukawa coupling $y$ (green lines) is chosen such that the correct DM relic abundance is achieved everywhere
below the DW line. Present and projected bounds from X-rays (filled gray
and black lines),
Lyman-$\alpha$ (orange), and DM self-interactions (violet)
are evaluated as described in the text. The two benchmark points \emph{BP1} and \emph{BP2} from Tab.~\ref{tab:BP},
see also Fig.~\ref{fig:evo}, are indicated as red stars.}
\label{fig:theta_ms}
\end{figure}

In Fig.~\ref{fig:theta_ms} we show a slice of the overall available parameter space for our setup in the $\sin^2(2\theta) - m_s$
plane, for a fixed mediator to DM mass ratio of  $m_\phi/m_s=3$.
For every point in parameter space, the dark sector Yukawa coupling $y$ is chosen such that the sterile neutrinos $\nu_s$ make
up all of DM after the era of exponential growth.
In the yellow band, DW production can give the correct relic abundance, including QCD and lepton flavor
uncertainties~\cite{Asaka:2006nq}; the dashed brown line corresponds to  the central prediction, which is the basis for our
choice of initial conditions for number and energy densities of $\nu_s$.
In the blue region DM will be overproduced, $\Omega_{s} h^2> 0.12$,
while the other filled regions are excluded by bounds
from X-ray searches (gray), Lyman-$\alpha$ observations (orange), and DM self-interactions (violet).
The white region corresponds to the presently allowed parameter space.

It is worth noting that, unlike in standard freeze-out scenarios~\cite{Bringmann:2009vf}, later kinetic decoupling
implies a {\it shorter} free-streaming length in our case
because the dark sector temperature scales as $T_{\rm d}\propto T_\nu^2$ already before that point. At the same time, the sound horizon 
increases for later kinetic decoupling. 
The shape of the Lyman-$\alpha$ exclusion lines reflects this, as kinetic decoupling occurs later for larger values of $y$. 

In Fig.~\ref{fig:theta_ms} we also show the projected sensitivities of the future X-ray experiments
eROSITA~\cite{Dekker:2021bos}, Athena~\cite{Ando:2021fhj}, and eXTP~\cite{Malyshev:2020hcc},
which will probe smaller values of $\sin^2(2\theta)$. Similarly, observables related to structure formation 
will likely result in improved future bounds, or in fact reveal anomalies that are not easily reconcilable with a standard 
non-interacting cold DM scenario.
While the precise reach is less clear here, we indicate with dashed orange and violet lines, respectively, 
the impact of choosing $\lambda_\mathrm{fs} < 0.12 \, \mathrm{Mpc}$, 
$r_\mathrm{s} < 0.15 \, \mathrm{Mpc}$, and $\sigma_T / m_s < 0.1 \, \mathrm{cm^2/g}$
rather than the corresponding limits described above.
Overall, prospects to probe a sizable region of the presently allowed parameter space 
appear very promising.

\smallskip
\paragraph*{Discussion.---}%
While an X-ray line would be the cleanest signature to claim DM discovery of the scenario suggested here,
let us briefly mention other possible directions. For example, the power-spectrum of DM density perturbations
at small, but only mildly non-linear, scales may be affected in a way that could be discriminated from alternative DM production
scenarios by 21\,cm and high-$z$ Lyman-$\alpha$ 
observations~\cite{Bose:2018juc,Munoz:2019hjh,Munoz:2020mue,Schaeffer:2021qwm}.
Another possibility would
be to search for a suppression of intense astrophysical neutrino fluxes due to $\phi$ production on $\nu_s$
DM at rest. We leave an investigation of these interesting avenues for future work.

We stress that the parameter space is larger than the $m_\phi/m_s=3$ slice shown in Fig.~\ref{fig:theta_ms}.
Larger mass ratios, in particular, have the effect of tightening (weakening)
bounds on $\lambda_\mathrm{fs}$ ($r_\mathrm{s}$), because kinetic decoupling happens earlier,
and weakening self-interaction constraints; this extends the viable parameter
space shown in Fig.~\ref{fig:theta_ms} to smaller mixing angles and allows for a larger range of $m_s$
(cf.\ Fig.~2 in the Appendix). 
Changing the interaction structure in the dark sector,
e.g.~by charging the sterile neutrinos under a gauge symmetry, is a further route for
model building that will not qualitatively change the new production scenario suggested here.

For completeness, we finally mention that smaller mediator masses are yet another,
though qualitatively different, route worthwhile to explore. For $m_s<m_\phi< 2m_s$
the mediator is no longer dominantly produced on-shell in transmission
processes, so the cross section for transmission,
$\nu_s\nu_\alpha\to\nu_s\nu_s$, scales as $y^4$ rather than $y^2$
and larger Yukawa couplings are needed in order to obtain the correct relic density. This, in turn,
implies that it may only be possible to satisfy the correspondingly tighter self-interaction and Lyman-$\alpha$ constraints
by adding a scalar potential for $\phi$ (because additional number-changing interactions would potentially allow an
increase in the $\nu_s$ abundance, similar to what
happens for the dashed green curve in Fig.~\ref{fig:evo}, right panel, at $m_s / T_\nu \sim 0.02$).
For even lighter mediators, $m_\phi<m_s$, extremely small Yukawa couplings or mixing angles
would be required to prevent DM from decaying too early through $\nu_s\to\nu_\alpha\phi$
(while the decay $\nu_s \to 3\nu_\alpha$, present also in the scenario we focus on here,
is automatically strongly suppressed as $\Gamma\propto y^4\sin^6\!\theta$).

\smallskip
\paragraph*{Conclusions.---}%
Sterile neutrinos constitute an excellent DM candidate. However,
X-ray observations rule out the possibility that these particles, in their simplest realization,
could make up all of the observed DM\@. On the other hand, there has been a
recent shift in focus in general DM theory, towards the possibility that DM may not just be a single, (almost)
non-interacting particle. Indeed, it is perfectly conceivable that DM could
belong to a more complex, secluded dark sector with its
own interactions and, possibly, further particles.

By combining these ideas in the most economic way, a sterile neutrino
coupled to a single additional dark sector degree of freedom allows for a qualitatively new DM production
mechanism and thereby opens up ample parameter space where sterile neutrinos could still explain the
entirety of DM\@. Excitingly, much of this parameter space is testable in the foreseeable future.
In particular, our results provide a strong motivation for further pushing
the sensitivity of X-ray line searches, beyond what would be expected from standard DW production.

\mbox{ }\\
\vfill
 \paragraph*{Acknowledgements.---}%
We would like to thank Hitoshi Murayama and Manibrata Sen for helpful discussions.  JTR also thanks Cristina Mondino, Maxim Pospelov, and Oren Slone for discussions on reproductive freeze-in applied to sterile neutrinos.
This work is supported by the Deutsche Forschungsgemeinschaft under
Germany's Excellence Strategy -- EXC 2121 `Quantum Universe'~-- 390833306,
the National Science Foundation under Grant No.\ NSF PHY-1748958,
and the National Research Foundation of Korea under Grant No.\ NRF-2019R1A2C3005009.
MH is further supported by the F.R.S./FNRS\@.  JTR is also supported by NSF grant PHY-1915409 and NSF CAREER grant
PHY-1554858. 
\newpage

\appendix

\section{Appendix}

Here we complement the discussion in the main text with additional, more  technical information
about the scenario that we study.

\begin{figure}[b]
	\includegraphics[width=\columnwidth]{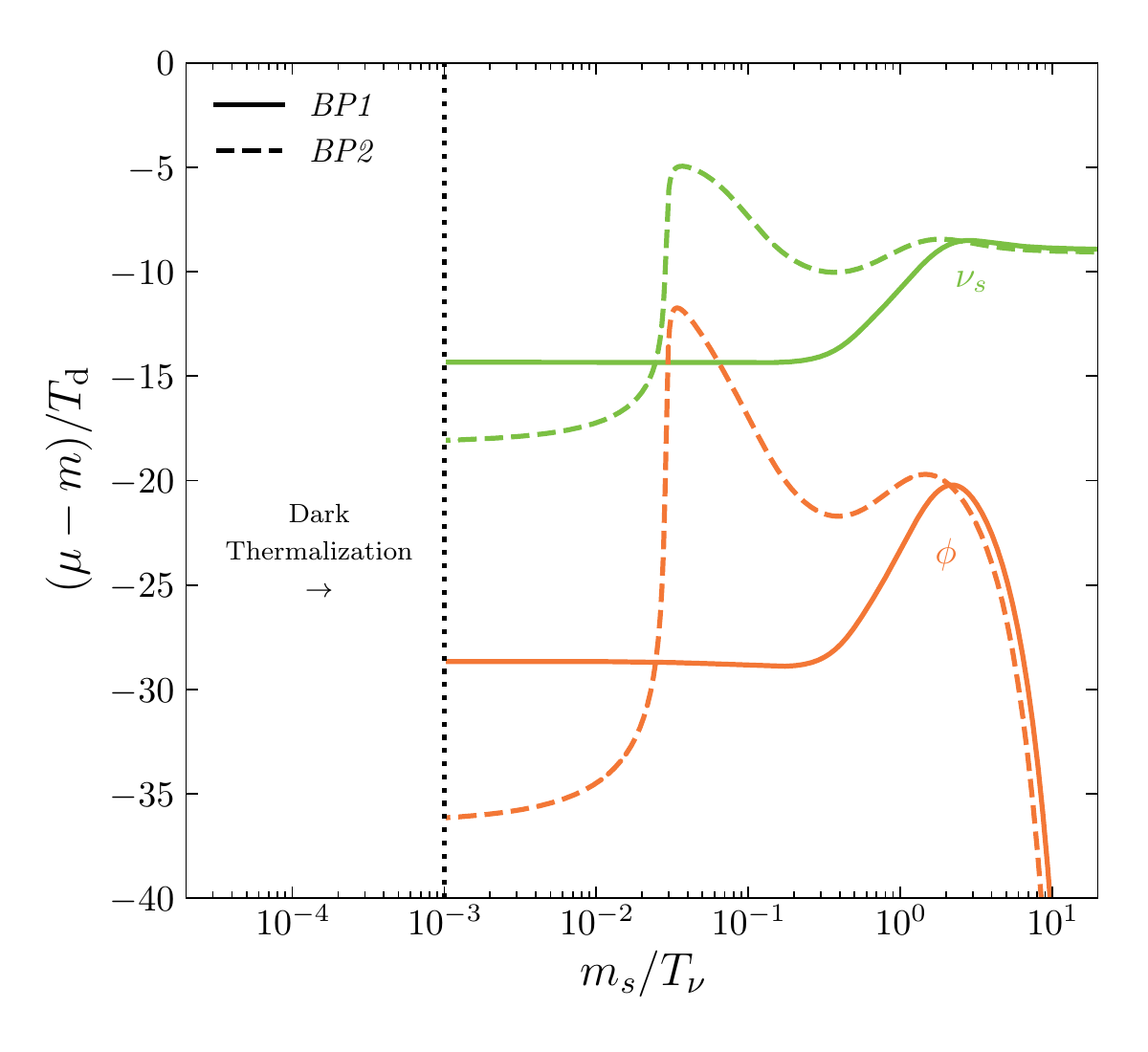}
	\caption{Evolution of the chemical potentials $\mu_s$ and $\mu_\phi$ for the benchmark points  \emph{BP1} (solid lines) and  
	\emph{BP2} (dashed lines) discussed in the main text.}
	\label{fig:evo_mu}
\end{figure}

\medskip
\paragraph*{Chemical potentials.---} In the main text, in Fig.~2, we showed and discussed the evolution of the 
dark sector temperature and number densities for the two benchmark points specified in Tab.~I. Here we complement 
in particular the discussion of the number densities by directly showing, in Fig.~\ref{fig:evo_mu}, the evolution of the chemical 
potentials  $\mu_s$ and $\mu_\phi$ for these benchmark points. In both cases, DW production leads to an abundance of sterile 
neutrinos with average momentum of the same order of magnitude as the SM
neutrino temperature at the time of production, but a highly suppressed
number density compared to the SM neutrino density. This leads to a
large negative chemical potential 
$\mu_s = \mu_\phi / 2 \ll - T_\mathrm{d}$ after thermalization in the dark sector. For \emph{BP1} (solid lines) this only changes 
when the exponential growth of the dark sector abundance starts.

For \emph{BP2} (dashed lines) the process $\nu_s \nu_s \to \phi \phi$ becomes important around $m_s/T_\nu \sim 0.01$, which 
increases the abundance by effectively converting $2 \nu_s \to 4 \nu_s$ (see main text). This transformation of kinetic energy 
to rest mass decreases the temperature and very efficiently increases the chemical potential. 
If equilibrium  with the inverse reaction $\phi \phi \to \nu_s \nu_s$ was to be established (enforcing $\mu_\phi = \mu_s$)
this would eventually result in a vanishing chemical potential (since $\mu_\phi = 2 \mu_s$ is still enforced by 
$\nu_s \nu_s \leftrightarrow \phi \phi$).  This point is never quite reached for \emph{BP2}, however, because
$\phi$ becomes Boltzmann suppressed due to $T_\mathrm{d} \ll m_\phi$ before that could happen.

\begin{figure}[t]
	\includegraphics[width=\columnwidth]{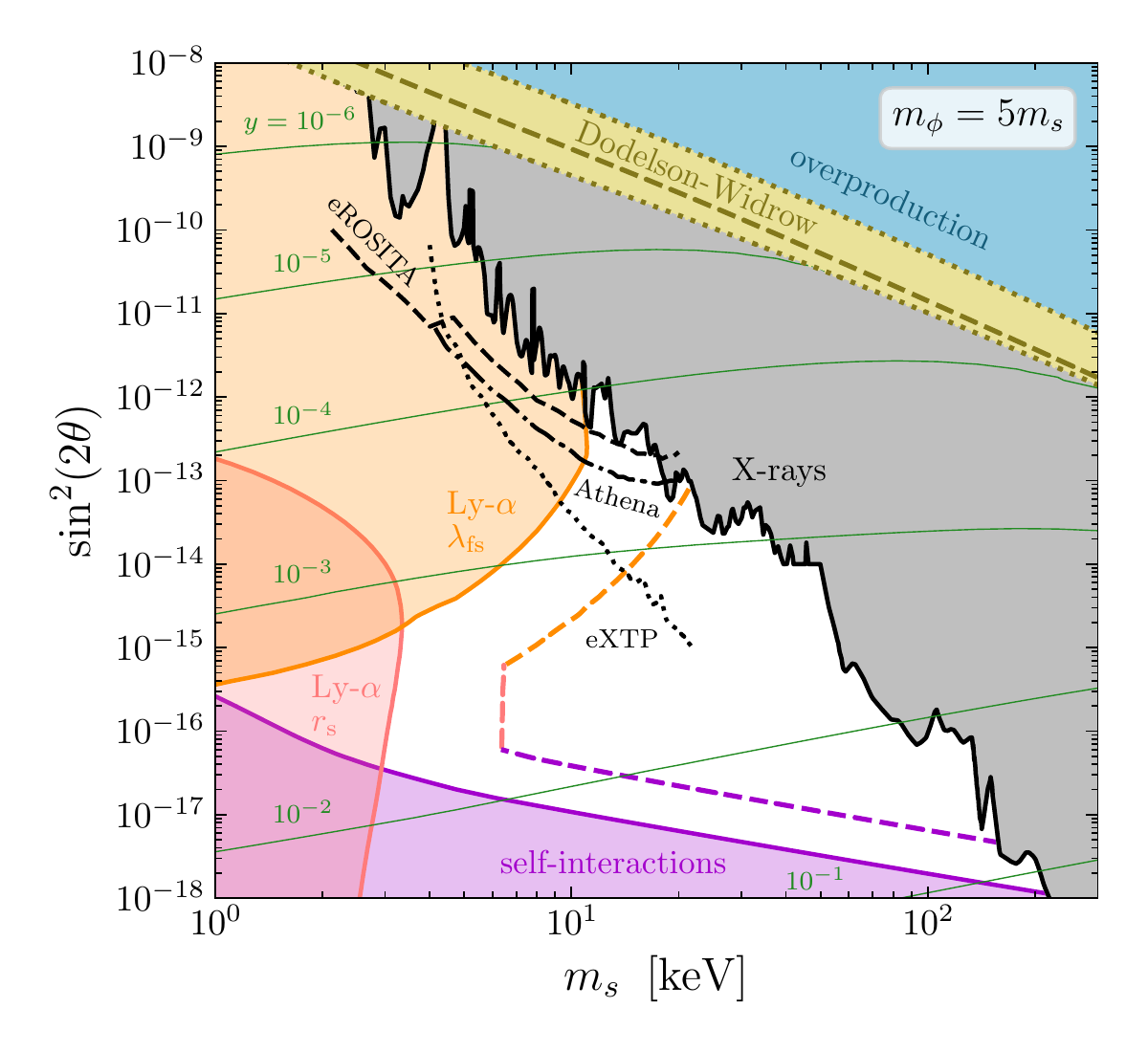}
	\caption{Same as Fig.~3 in the main text, but for $m_\phi = 5 m_s$.}
	\label{fig:theta_ms_ratio_5}
\end{figure}

\medskip
\paragraph*{Other mass ratios.---}
In the main text, we already discussed the general impact of changing the mass ratio of  $m_\phi / m_s=3$ that
we chose for defining our benchmark points, as well as for illustrating the newly opened 
parameter space for sterile neutrino DM in Fig.~3 (of the main text). In Fig.~\ref{fig:theta_ms_ratio_5} we further complement
this discussion by explicitly showing the situation for $m_\phi / m_s=5$. We note that, apart from the general
strengthening or weakening of constraints 
discussed in the main text, the 
Lyman-$\alpha$ forest constraint on the sound horizon first becomes stronger when increasing from very small mixing angles, 
against what would be expected from an earlier kinetic decoupling due to a smaller Yukawa $y$. This feature can be explained by the 
fact that at larger $y$ also the process $\nu_s \nu_s \to \phi \phi$ is more important, which can decrease $T_\mathrm{d}$ (after the 
decays of $\phi$) and thereby lower the speed of sound.

\medskip
\paragraph*{Collision operators.---}
Here, we provide explicit expressions for the types of collision operators that we used in our analysis.
As discussed in the main text, we can restrict ourselves to $1\to2$, $2\to1$, and $2\to2$ interactions. 
We always state the collision operators $C_n$ for the number density of some particle $\chi \in \{ \nu_s, \, \phi \}$ from a given reaction, assuming that $\chi$ appears only once in the initial state (for an appearance in the final 
state the expressions need to be multiplied by $-1$). Multiple occurrences can be treated by summation over the corresponding 
expressions, and the total collision operator is given by the sum over all possible processes. 
We do not state explicitly the collision operators $C_\rho$ for the energy density; these can be obtained in the same way
as the $C_n$, after multiplying the integrand with the energy $E_\chi$ of $\chi$.

We assume $CP$-conservation for all reactions and denote particles by numbers $i\in\{1,2,3,4\}$ ($\chi$ being one of the 
particles $i$), four-momenta by $\ul{p}_i$, three-momenta by $\bold{p}_i$, absolute values of three-momenta by $p_i$, 
and energies by $E_i = \sqrt{m_i^2 + p_i^2}$, where $m_i$ is the particle's mass. 
The matrix elements in our expressions are summed over the initial- and final-state spin degrees of freedom.
As the dark sector is in general not non-relativistic (or satisfies  $\mu \ll -T_\mathrm{d}$), we fully keep spin-statistical factors 
and the complete Fermi-Dirac or Bose-Einstein distributions everywhere. This is particularly important for scenarios 
where the reaction $\nu_s \nu_s \to \phi \phi$ is relevant (roughly $y \gtrsim 4 \times 10^{-4}$ in Fig.~3 in the main text, cf.~dashed 
lines in main text Fig.~2) and can lead to a change in the final sterile neutrino abundance by more than two orders of magnitude
compared to a simplified treatment assuming Boltzmann distributions and no Pauli blocking or Bose enhancement.\footnote{Due to 
the exponential dependence on $y$, this however only changes the required value of $y$ for the correct DM abundance 
by a few percent.}

Starting with inverse decays, $12\to 3$, the collision operator for the number density takes the form
\begin{align}
\frac{C_{n, 1 2 \to 3}}{ \kappa_{12 \to 3}} &=  \int \frac{\d^3 p_1}{(2 \pi)^3 2 E_1} \frac{\d^3 p_2}{(2 \pi)^3 2 E_2} \frac{\d^3 p_3}{(2 \pi)^3 2 E_3} (2 \pi)^4 \nonumber \\
 &\quad \; \times \delta ( \ul{p}_1 + \ul{p}_2 - \ul{p}_3) f_1 f_2 (1 \pm f_3) | \mathcal{M}_{1 2 \to 3} |^2 \nonumber \\
&= \frac{1}{4 (2 \pi)^3} \int_{m_1}^\infty \d E_1 \int_{R_2} \d E_3 \nonumber \\
&\quad \; \times f_1 f_2 (1 \pm f_3)|_{E_3 = E_1 + E_2} | \mathcal{M}_{1 2 \to 3} |^2\eqsp,\\
R_2 &= \{ E_2 \geq m_2 \; | \; E_{2,-} \leq E_2 \leq E_{2,+} \} \eqsp, \\
E_{2, \pm} &= \frac{1}{2 m_1^2} \Bigg( E_1 [m_3^2 - m_1^2 - m_2^2] \\
&\quad \;  \pm p_1 \sqrt{m_3^4 + (m_1^2 \!-\! m_2^2)^2 - 2 m_3^2 (m_1^2 \!+\! m_2^2)} \Bigg)\eqsp, \nonumber
\end{align}
where the symmetry factor $\kappa_{12 \to 3} = 1/2$ if particles 1 and 2 are identical and $\kappa_{12 \to 3} = 1$ if not, and the 
phase-space distribution functions are denoted by the $f_i$ (with $\pm$ denoting Bose enhancement/Pauli suppression if $3$ is a 
boson/fermion, analogously for other particles henceforth).
Similarly, for the reverse reaction $3 \to 1 2$, we find
\begin{align}
\frac{C_{n, 3 \to 1 2}}{\kappa_{3 \to 12}} &= \frac{1}{4 (2 \pi)^3} \int_{m_1}^\infty \d E_1 \int_{R_2} \d E_3 \nonumber \\
&\times f_3 (1 \pm f_2) (1 \pm f_1)|_{E_3 = E_1 + E_2} | \mathcal{M}_{1 2 \to 3} |^2\eqsp.
\end{align}

Turning to 2-to-2 reactions, with particle identifications $1 2 \to 3 4$, the collision operator for the number density is given by
\begin{align}
\frac{C_{n,1 2 \to 3 4}}{\kappa_{12 \to 34}} = &\int \frac{\d^3 p_1}{(2 \pi)^3 2 E_1} \frac{\d^3 p_2}{(2 \pi)^3 2 E_2} \frac{\d^3 p_3}{(2 \pi)^3 2 E_3} \frac{\d^3 p_4}{(2 \pi)^3 2 E_4} \nonumber \\
&\times (2 \pi)^4 \delta (\ul{p}_1 + \ul{p}_2 - \ul{p}_3- \ul{p}_4) \nonumber \\
&\times f_1 f_2 (1 \pm f_3) (1 \pm f_4) | \mathcal{M}_{1 2 \to 3 4} |^2\,,
\end{align}
with the symmetry factor $\kappa_{12 \to 34}$ being 1 if neither the initial-state nor the final-state particles are identical, $1/2$ if 
either the initial-state or the final-state particles are identical, and $1/4$ if the initial-state and the final-state particles are identical. 
We follow a similar approach as discussed in Ref.~\cite{Ala-Mattinen:2022nuj} for the Boltzmann equation at the phase-space 
level. After integrating over $\bold{p}_4$ using the spatial part of the $\delta$-distribution, we can write the 3-momenta 
in spherical coordinates as
\begin{align}
\bold{p}_1 &= p_1 (0, 0, 1)^T\eqsp, \\
\bold{p}_2 &= p_2 (\sin \beta, 0, \cos \beta)^T \\
\bold{p}_3 &= p_3 (\sin \theta \cos \phi, \sin \theta \sin \phi, \cos \theta)^T\eqsp,
\end{align}
where $-1 \leq \cos \beta \leq 1$, $0 \leq \phi < 2 \pi$, and $- 1\leq \cos \theta \leq 1$, 
and hence
\begin{align}
\bold{p}_4^2 &= (\bold{p}_1 + \bold{p}_2 - \bold{p}_3)^2 \nonumber \\
&= p_1^2 + p_2^2 + p_3^2 + 2 p_1 p_2 \cos \beta - 2 p_1 p_3 \cos \theta \nonumber \\
&\quad - 2 p_2 p_3 (\cos \beta \cos \theta + \sin \beta \sin \theta \cos \phi)\eqsp.
\end{align}
Since the $\phi$-dependence of the entire integrand is only through $\cos \phi$, we can simply multiply by 2 and restrict the 
integration to $0 \leq \phi \leq \pi$. In particular, the matrix element only depends on the Mandelstam variables 
\begin{align}
s &= m_1^2 + m_2^2 +2 E_1 E_2 - 2 p_1 p_2 \cos \beta\eqsp, \label{eq:man_s} \\
t &= m_1^2 + m_3^2 -2 E_1 E_3 + 2 p_1 p_3 \cos \theta\,. \label{eq:man_t} 
\end{align}
As a result, we can write the remaining $\delta$-distribution as
\begin{align}
&\delta (E_1 + E_2 - E_3 - E_4|_{\bold{p}_4 = \bold{p}_1 + \bold{p}_2 - \bold{p}_3}) = \nonumber \\
&\frac{E_1 + E_2 - E_3}{p_2 p_3 \sin \beta \sin \theta}\delta \Big( \cos \phi \nonumber \\
&- \frac{1}{2 p_2 p_3 \sin \beta \sin \theta} [m_1^2 + m_2^2 + m_3^2 - m_4^2 \nonumber \\
&+ 2 (E_1 E_2 - E_1 E_3 - E_2 E_3) \nonumber \\
&- 2 p_1 p_2 \cos \beta + 2 p_1 p_3 \cos \theta + 2 p_2 p_3 \cos \beta \cos \theta] \Big)\eqsp.
\end{align}
The integration over $\cos \theta$ gets restricted to $\cos \theta \in R_\theta$ with $R_\theta = \{-1 \leq \cos \theta \leq 1 \; | \; c_{\theta, +} \leq \cos \theta \leq c_{\theta, -} \}$ and
\begin{align}
c_{\theta, \pm} &= \frac{-b \pm \sqrt{b^2-4 a c}}{2 a}\eqsp, \\
a &= -4 p_3^2 [(E_1 + E_2)^2 - s]\eqsp, \\
b &= \frac{2 p_3}{p_1} [s - 2 E_1 (E_1+E_2) + m_1^2 - m_2^2] \nonumber \\
&\quad \; \times [s - 2 E_3 (E_1+E_2) + m_3^2 - m_4^2]\eqsp,\\
c &= -[2 E_3 (E_1 + E_2)-m_3^2+m_4^2-s]^2 \nonumber \\
&\quad \; - \frac{p_3^2}{p_1^2} (s - s_{12,-}) (s - s_{12,+})\,, \\
s_{12/34, \pm} &= m_{1/3}^2 + m_{2/4}^2 + 2 E_{1/3} E_{2/4} \pm 2 p_{1/3} p_{2/4}\eqsp,\label{eq:b24ac}
\end{align}
with $E_4 = E_1 + E_2 - E_3$ and $p_4 = \sqrt{E_4^2 - m_4^2}$. Performing the integration over $\phi$ we  thus obtain
\begin{align}
&\quad \;  \int \frac{\d^3 p_3}{(2 \pi)^3 2 E_3} \frac{\d^3 p_4}{(2 \pi)^3 2 E_4} (2 \pi)^4 \delta (\ul{p}_1 + \ul{p}_2 - \ul{p}_3- \ul{p}_4) \nonumber  \\ 
&\quad \;   \times (1 \pm f_3) (1 \pm f_4) | \mathcal{M}_{1 2 \to 3 4} |^2  \\
&=\frac{1}{(2 \pi)^2} \int_0^\infty \d p_3 \frac{p_3^2}{E_3} (1 \pm f_3) (1 \pm f_4) \theta (b^2-4 a c) \nonumber\\
&\quad \;   \times \int_{R_\theta} \frac{\d \cos \theta}{\sqrt{a \cos^2 \theta + b \cos \theta + c}} | \mathcal{M}_{1 2 \to 3 4} |^2\eqsp,
\end{align}
where we used $2 p_2 p_3 \sin \beta \sin \theta \sqrt{1-\cos^2 \phi} = \sqrt{a \cos^2 \theta + b \cos \theta + c}$
and introduced a factor of $\theta (b^2-4 a c)$ to ensure that $c_{\theta, \pm}$ is real. 
After variable transformation from $\cos \beta$ to $s$, cf.~Eq.~(\ref{eq:man_s}), we thus finally arrive at
\begin{align}
&C_{n,1 2 \to 3 4} = \frac{\kappa_{12 \to 34}}{4 (2 \pi)^6} \int_{m_1}^{\infty} \d E_1 \int_{\max (m_2, m_3+m_4-E_1)}^\infty \d E_2 \nonumber \\
&\quad \times \int_{m_3}^{E_1+E_2-m_4} \d E_3 \, p_3 f_1 f_2 (1 \pm f_3) (1 \pm f_4) \nonumber \\
&\quad \times \int_{R_s} \d s \int_{R_\theta} \frac{\d \cos \theta}{\sqrt{a \cos^2 \theta + b \cos \theta + c}} | \mathcal{M}_{1 2 \to 3 4} |^2\eqsp,
\end{align}
where $R_s = \{ s \in \mathbb{R} \; | \; \max (s_{12,-}, s_{34,-}) \leq s \leq \min (s_{12,+}, s_{34,+}) \}$
and  the integration over $\cos \theta$ can equally well be rewritten as over $t$, cf.~Eq~(\ref{eq:man_t}).
We can also rearrange the integration order to
\begin{align}
C_{n,1 2 \to 3 4} &= \frac{\kappa_{12 \to 34}}{4 (2 \pi)^6} \int_{s_\mathrm{min}}^\infty \d s \int_{m_1}^{\infty} \d E_1 \int_{R_2} \d E_2 \int_{R_3} \d E_3 \nonumber \\
&\quad\;\times p_3 f_1 f_2 (1 \pm f_3) (1 \pm f_4) \nonumber \\
&\quad\;\times \int_{R_\theta} \frac{\d \cos \theta}{\sqrt{a \cos^2 \theta + b \cos \theta + c}} | \mathcal{M}_{1 2 \to 3 4} |^2\eqsp,
\end{align}
\newpage
where
\begin{align}
s_\mathrm{min} &= \max ( [m_1+m_2]^2, [m_3 + m_4]^2 )\eqsp, \\
R_2 &= \{ E_2 \geq m_2 \; | \; E_{2,-} \leq E_2 \leq E_{2,+} \}\eqsp, \\
R_3 &= \{ E_3 \geq m_3 \; | \; E_{3,-} \leq E_3 \leq E_{3,+} \}\eqsp, \\
E_{2,\pm} &= \frac{1}{2 m_1^2} \Bigg(E_1 [s-m_1^2-m_2^2] \nonumber \\
&\quad \; \pm p_1 \sqrt{s^2 + (m_1^2 - m_2^2)^2 - 2 s (m_1^2 + m_2^2)} \Bigg)\eqsp, \\
E_{3,\pm} &= \frac{1}{2 s} \Bigg([E_1+E_2] [s+m_3^2-m_4^2] \pm \sqrt{(E_1+E_2)^2-s} \nonumber \\
& \quad\;\times \sqrt{s^2 + (m_3^2 - m_4^2)^2 - 2 s (m_3^2 + m_4^2)} \Bigg)\eqsp.
\end{align}
For the matrix elements of interest in this work, see below, the integration over $\cos \theta$ can be performed analytically 
after a variable transformation to $x = -2 \arcsin (\sqrt{(c_{\theta, -} - \cos \theta)/(c_{\theta, -} - c_{\theta, +})} )$. 
The remaining four integrals we evaluate numerically, using Monte-Carlo methods with importance sampling.

\medskip
\paragraph*{Matrix elements.---}
For completeness, we finally provide a full list of all matrix elements that are relevant for our scenario (summed over the spin 
degrees of freedom of all initial- and final-state particles). Note that we can safely assume $CP$-conservation for these. We start 
with the  decay width
\be
\Gamma_\phi = \Gamma_{\phi \to \nu_\alpha \nu_\alpha} + \Gamma_{\phi \to \nu_\alpha \nu_s} + \Gamma_{\phi \to \nu_s \nu_s}
\ee
 of the mediator $\phi$, implicitly defining the matrix elements via the partial widths
\begin{align}
\Gamma_{\phi \to \nu_\alpha \nu_\alpha} &= \frac{1}{32 \pi m_\phi} |\mathcal{M}_{\phi \to \nu_\alpha \nu_\alpha}|^2 = \frac{1}{8 \pi} y^2 \sin^4\!\theta \, m_\phi \eqsp,
\end{align}
\begin{align}
\Gamma_{\phi \to \nu_\alpha \nu_s} &= \frac{m_\phi^2 - m_s^2}{16 \pi m_\phi^3} |\mathcal{M}_{\phi \to \nu_\alpha \nu_s}|^2 \Theta (m_\phi - m_s)  \\
&= \frac{1}{8 \pi} y^2 \sin^2 \theta \cos^2 \theta \, \frac{(m_\phi^2 - m_s^2)^2}{m_\phi^3} \Theta (m_\phi - m_s) \eqsp, \nonumber
\end{align}
\begin{align}
\Gamma_{\phi \to \nu_s \nu_s} &= \frac{(m_\phi^2 - 4 m_s^2)^{1/2}}{32 \pi m_\phi^2} |\mathcal{M}_{\phi \to \nu_s \nu_s}|^2 \Theta (m_\phi - 2 m_s) \nonumber \\
&= \frac{1}{8 \pi} y^2 \cos^4 \theta \, \frac{(m_\phi^2 - 4 m_s^2)^{3/2}}{m_\phi^2} \Theta (m_\phi - 2m_s)\eqsp.
\end{align}
We note that, typically, $\Gamma_\phi \simeq \Gamma_{\phi \to \nu_s \nu_s}$ as $\sin \theta \ll 1$.

\begin{widetext}
Turning to $2\to2$ processes, we find
\begin{align}
|\mathcal{M}_{\nu_s \nu_s \to \phi \phi}|^2 = 2 y^4 \cos^8 \theta \times \frac{(s+2 t-2 m_\phi^2 -2 m_s^2) (-m_s^4-m_\phi^4-m_s (2 m_\phi^2-s-2 t)+2 m_\phi^2 t - t (s+t))}{(t-m_s^2)^2 (u-m_s^2)^2}\eqsp.
\end{align}
for the matrix element of $\nu_s \nu_s \to \phi \phi$, and
for $\nu_1 \nu_2 \to \nu_3 \nu_4$ -- with 1, 2, 3, and 4 being $s$ or $\alpha$ -- we have
\begin{align}
&|\mathcal{M}_{\nu_1 \nu_2 \to \nu_3 \nu_4}|^2 = 4 \tilde{y}^2 \Bigg[ \frac{((m_1 + m_2)^2-s) ((m_3 + m_4)^2-s)}{(s-m_\phi^2)^2 + m_\phi^2 \Gamma_\phi^2} + \frac{((m_1 + m_3)^2-t) ((m_2 + m_4)^2-t)}{(t-m_\phi^2)^2 + m_\phi^2 \Gamma_\phi^2} \nonumber \\
&+ \frac{((m_1 + m_4)^2-u) ((m_2 + m_3)^2-u)}{(u-m_\phi^2)^2 + m_\phi^2 \Gamma_\phi^2} - \frac{(s-m_\phi^2) (t-m_\phi^2) + m_\phi^2 \Gamma_\phi^2}{[(s-m_\phi^2)^2 + m_\phi^2 \Gamma_\phi^2] [(t-m_\phi^2)^2 + m_\phi^2 \Gamma_\phi^2]} \nonumber \\
&\times [m_1^3 m_4 + m_1^2 m_4 (m_2+m_3+m_4) + m_2 m_3 (m_2^2 + (m_2+m_3) (m_3 + m_4)) - m_3 (m_2+m_4) s- (m_2 (m_3+m_4)+s) t \nonumber \\
&+ m_1 (m_2^2 m_3 + m_4 (m_4 (m_3+m_4) -s)  + m_2 ((m_3+m_4)^2-s)-(m_3+m_4) t)] \nonumber \\
&-\frac{(s-m_\phi^2) (u-m_\phi^2) + m_\phi^2 \Gamma_\phi^2}{[(s-m_\phi^2)^2 + m_\phi^2 \Gamma_\phi^2] [(u-m_\phi^2)^2 + m_\phi^2 \Gamma_\phi^2]} [m_1^3 m_3+m_1^2 m_3 (m_2+m_3+m_4) + m_2 m_4 (m_2^2 + (m_2 + m_4) (m_3+m_4))\nonumber \\
&- (m_2+m_3) m_4 s - (m_2 (m_3+m_4) + s) u \nonumber \\
&+ m_1 (m_2^2 m_4 + m_3 (m_3 (m_3+m_4)-s) + m_2 ((m_3+m_4)^2-s) - (m_3+m_4) u)] \nonumber \\
&-\frac{(t-m_\phi^2) (u-m_\phi^2) + m_\phi^2 \Gamma_\phi^2}{[(t-m_\phi^2)^2 + m_\phi^2 \Gamma_\phi^2] [(u-m_\phi^2)^2 + m_\phi^2 \Gamma_\phi^2]} [m_1^3 m_2 + m_1^2 m_2 (m_2+m_3+m_4) + m_3 m_4 (m_4^2 +(m_2 + m_3) (m_3+m_4)) \nonumber \\
&- m_4 (m_2+m_3) t - (m_3 (m_2+m_4)+t) u \nonumber \\
&+ m_1 (m_2^3 + m_2^2 (m_3+m_4) +m_3 m_4 (m_3+m_4) - m_3 t + m_2 (2 m_3 m_4 - t - u) - m_4 u)] \Bigg]\,. 
\label{eq:M1234_scalar}
\end{align}
\end{widetext}
Here the Mandelstam variables $s$, $t$, and $u$ follow the standard definition, and $\tilde{y}^2 = y^4 \cos^8 \theta$ for elastic 
scattering $\nu_s \nu_s \to \nu_s \nu_s$, $\tilde{y}^2 = y^4 \cos^4 \theta \sin^4 \theta$ for the transformation process 
$\nu_s \nu_\alpha \to \nu_s \nu_s$, and $\tilde{y}^2 = y^4 \cos^4 \theta \sin^4 \theta$ for freeze-in 
$\nu_\alpha \nu_\alpha \to \nu_s \nu_s$ (which can generally be neglected). 

In the main text, we have stressed that dark sector thermalization and the phase of exponential growth are mostly due to the 
3-body interactions $\nu_s \nu_s \leftrightarrow \phi$ and $\nu_s \nu_\alpha \to \phi$. These scale as $y^2$ and are equal to the 
on-shell contribution to the $s$-channel part of $\nu_1 \nu_2 \to \nu_3 \nu_4$, cf.~the first term in Eq.~(\ref{eq:M1234_scalar}).
From the full expression of the matrix element, see also Fig.~1 in the main text, it is however evident that there are also
contributions 
from $t/u$-channel diagrams, as well as $s$-channel contributions with an off-shell mediator.
The processes $\nu_s \nu_s \to \nu_s \nu_s$ and $\nu_s \nu_\alpha \to \nu_s \nu_s$ thus have 
interaction rates scaling as $y^4$ far away from the $s$-channel resonance. 
Since $\nu_s \nu_\alpha \to \phi$ is at most barely larger than the Hubble rate (cf.\ left panel of Fig.~2 in the main text), 
the additional suppression by $y^2$ causes the off-shell contribution to $\nu_s \nu_\alpha \to \nu_s \nu_s$ to be negligible.

As also discussed in the main text, it is sufficient to solve the Boltzmann equations for $\tilde{n} = n_s + 2 n_\phi$ and 
$\rho = \rho_s + \rho_\phi$ since $\nu_s \nu_s \leftrightarrow \phi$ thermalizes the dark sector. Self-scatterings 
$\nu_s \nu_s \leftrightarrow \nu_s \nu_s$ do not change number or energy density, i.e.~their contribution to the corresponding 
collision operators is zero.
Similarly,  $\nu_s \nu_s \leftrightarrow \phi$ does not change $\tilde{n}$ or $\rho$ and does not need to be calculated for the 
evolution. Note however that the full process $\nu_s \nu_s \to \nu_s \nu_s$ \emph{is} relevant for kinetic decoupling, which can 
occur after $\phi$ becomes highly Boltzmann suppressed at $T_\mathrm{d} \ll m_\phi$.

\bibliography{pandemic.bib}

\end{document}